\documentclass{article}

\usepackage{arxiv}

\usepackage[utf8]{inputenc} 
\usepackage[T1]{fontenc}    
\usepackage{lmodern}        
\usepackage{hyperref}       
\usepackage{url}            
\usepackage{booktabs}       
\usepackage{amsfonts}       
\usepackage{nicefrac}       
\usepackage{microtype}      
\usepackage{graphicx}

\title{Empirical Evidence That There Is No Such Thing As A Validated
Prediction Model}

\author{
    Florian D. van Leeuwen
   \\
    Biomedical Data Sciences \\
    Leiden University Medical Center \\
  Leiden, The Netherlands \\
  \texttt{\href{mailto:f.d.van_leeuwen@lumc.nl}{\nolinkurl{f.d.van\_leeuwen@lumc.nl}}} \\
   \And
    Ewout W. Steyerberg
   \\
    Biomedical Data Sciences \\
    Leiden University Medical Center \\
  Leiden, The Netherlands \\
  \texttt{\href{mailto:e.w.steyerberg@lumc.nl}{\nolinkurl{e.w.steyerberg@lumc.nl}}} \\
   \And
    David van Klaveren
   \\
    Public Health \\
    Erasmus University Medical Center \\
  Rotterdam, The Netherlands \\
  \texttt{\href{mailto:d.vanklaveren@erasmusmc.nl}{\nolinkurl{d.vanklaveren@erasmusmc.nl}}} \\
   \And
    Ben Wessler
   \\
    Predictive Analytics and Comparative Effectiveness Center \\
    Tufts Medical Center \\
  Boston, MA, USA \\
  \texttt{\href{mailto:Benjamin.S.Wessler@tuftsmedicine.org}{\nolinkurl{Benjamin.S.Wessler@tuftsmedicine.org}}} \\
   \And
    David M. Kent
   \\
    Predictive Analytics and Comparative Effectiveness Center \\
    Tufts Medical Center \\
  Boston, MA, USA \\
  \texttt{\href{mailto:David.Kent@tuftsmedicine.org}{\nolinkurl{David.Kent@tuftsmedicine.org}}} \\
   \And
    Erik W. van Zwet
   \\
    Biomedical Data Sciences \\
    Leiden University Medical Center \\
  Leiden, The Netherlands \\
  \texttt{\href{mailto:e.w.van_zwet@lumc.nl}{\nolinkurl{e.w.van\_zwet@lumc.nl}}} \\
  }

\newlength{\cslhangindent}
\newlength{\csllabelwidth}
\newlength{\cslentryspacingunit} 
  {}%
  {\par}
\newenvironment{CSLReferences}[2] 
 {
  \setlength{\parindent}{0pt}
  \ifodd #1
  \let\oldpar\par
  \def\par{\hangindent=\cslhangindent\oldpar}
  \fi
  \setlength{\parskip}{#2\cslentryspacingunit}
 }%
 {}
\usepackage{calc}

\usepackage{amsmath}
\begin{document}
\maketitle

\begin{abstract}
\textbf{Background} External validations are essential to assess the
performance of a clinical prediction model (CPM) before deployment.
Apart from model misspecification, also differences in patient
population, standard of care, predictor definitions and other factors
influence a model's discriminative ability, as commonly quantified by
the AUC (or c-statistic). We aimed to quantify the variation in AUCs
across sets of external validation studies, and propose ways to adjust
expectations of a model's performance in a new setting.

\textbf{Methods} The Tufts-PACE CPM Registry holds a collection of CPMs
for prognosis in cardiovascular disease. We analyzed the AUC estimates
of 469 CPMs with at least one external validation. Combined, these CPMs
had a total of 1,603 external validations reported in the literature.
For each CPM and its associated set of validation studies, we performed
a random effects meta-analysis to estimate the between-study standard
deviation \(\tau\) among the AUCs. Since the majority of these
meta-analyses has only a handful of validations, this leads to very poor
estimates of \(\tau\). So, instead of focusing on a single CPM, we
estimated a log normal distribution of \(\tau\) across all 469 CPMs. We
then used this distribution as an empirical prior. We used
cross-validation to compare this empirical Bayesian approach with
frequentist fixed and random effects meta-analyses.

\textbf{Results} The 469 CMPs included in our study had a median of 2
external validations with an IQR of {[}1-3{]}. The estimated
distribution of \(\tau\) had mean 0.055 and standard deviation 0.015. If
\(\tau\) = 0.05, then the 95\% prediction interval for the AUC in a new
setting has a width of at least +/- 0.1, no matter how many validations
have been done. The usual frequentist methods grossly underestimate the
uncertainty about the AUC in a new setting. Accounting for \(\tau\) in a
Bayesian approach achieved near nominal coverage.

\textbf{Conclusion} Due to large heterogeneity among the validated AUC
values of a CPM, there is great irreducible uncertainty in predicting
the AUC in a new setting. This uncertainty is underestimated by existing
methods. The proposed empirical Bayes approach addresses this problem
which merits wide application in judging the validity of prediction
models.
\end{abstract}


\hypertarget{introduction}{%
\section{Introduction}\label{introduction}}

Clinical prediction models may provide care-givers and patients with
quantitative estimates of risk and prognosis, which can inform clinical
decision-making (E. W. Steyerberg 2009). Before deployment of a newly
developed CPM, it is crucial that its performance is carefully and
repeatedly validated. If the performance of a CPM is assessed with the
same data that was used to develop it, then it is important account for
some degree of overfitting. Common approaches for internal validation
include cross-validation and bootstrap resampling (Harrell 2015). Beyond
internal validation, external validation refers to the assessment of
performance in a new setting (a plausibly related population (Justice
1999)). While internal validation quantifies reproducibility, external
validation assesses the generalizability of CPMs (Altman and Royston
2000; Justice 1999; Ewout W. Steyerberg and Harrell 2016).

Here we study the Tufts-PACE CPM Registry which is a unique, carefully
curated set of external validations of CPMs in the field of
cardiovascular medicine (Wessler et al. 2021). We focus on
discrimination as a key aspect of performance at external validation
studies, commonly quantified in terms of the Area Under the Receiver
Operating Curve (AUROC, AUC) or the c-statistic. Large variation among
the validations of the same CPM would problematic because it implies
that there is great uncertainty about the AUC when we want to deploy
that CPM in a new setting. Therefore, our main goal is to assess the
amount of heterogeneity among the validations of a CPM and propose ways
to adjust expectations of a model's performance in a new setting.
Moreover, as we will demonstrate, the usual frequentist methods severely
underestimate this uncertainty.

The paper is organized as follows. In the next section we introduce our
data set, provide the relevant background information and introduce the
problem with two examples. In section 3 we describe our statistical
model, and propose an empirical Bayes approach for predicting the AUC in
a new setting. In section 4 we present our results. We provide an
estimate of the heterogeneity and use cross-validation to compare our
empirical Bayes approach to the usual (frequentist) methods. We end the
paper with a brief discussion.

\hypertarget{background-and-problem-statement}{%
\section{Background and problem
statement}\label{background-and-problem-statement}}

\begin{figure}[ht]

{\centering \includegraphics[width=0.55\linewidth]{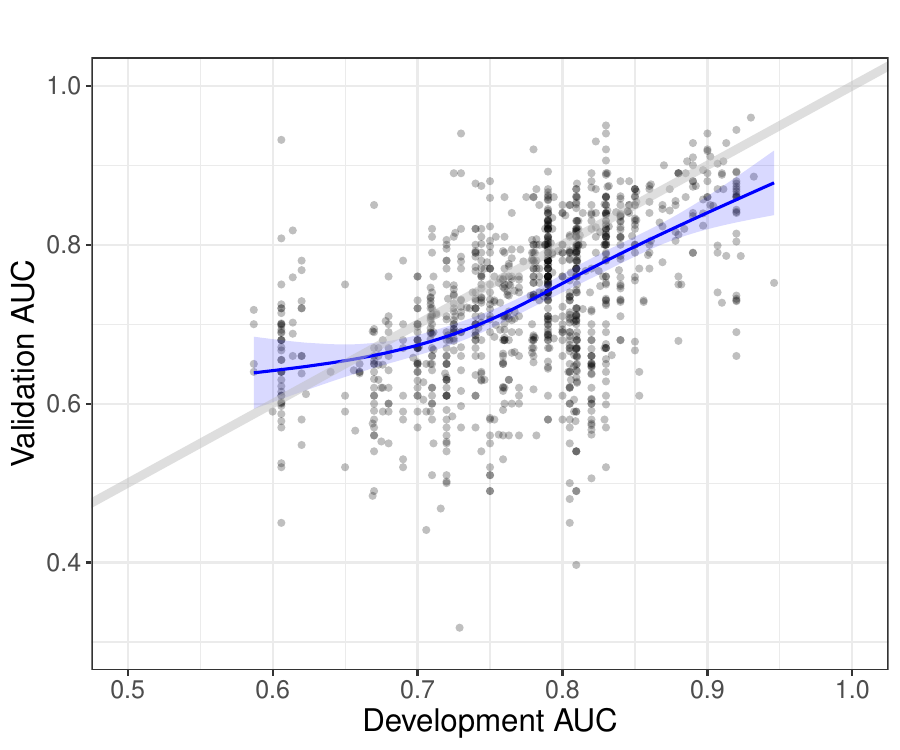} 

}

\caption{Relation between development AUCs and validation AUCs in the Tufts-PACE CPM Registry. The regression curve shows that the validation AUCs tend to be lower than the development AUCs.\label{fig:1}}\label{fig:unnamed-chunk-2}
\end{figure}

As an introduction to our data set, we plot the external validation AUCs
(or c-statistics) versus the associated development AUCs (Figure
\ref{fig:1}). We added a regression curve (a natural spline with 3
degrees of freedom) and note that the AUCs at development were
systematically higher than AUCs at validation. This may be due to
optimism that is not always fully accounted for at internal validation.
Moreover, validation populations may be more or less heterogeneous than
the development population. We also note a substantial variability
across validation AUCs.

As an example, we consider the CRUSADE prediction model for patients
with angina pectoris (Subherwal et al. 2009). This model was externally
validated one year after development (Abu-Assi et al. 2010). The
external validation resulted in an estimated AUC of 0.82 with 95\%
confidence interval from 0.77 to 0.87. This would seem to imply that if
we would use this CPM in a new setting, we can be quite confident that
the AUC will be at least 0.77. Unfortunately, that is not the case at
all.

\begin{figure}[ht]

{\centering \includegraphics[width=0.7\linewidth]{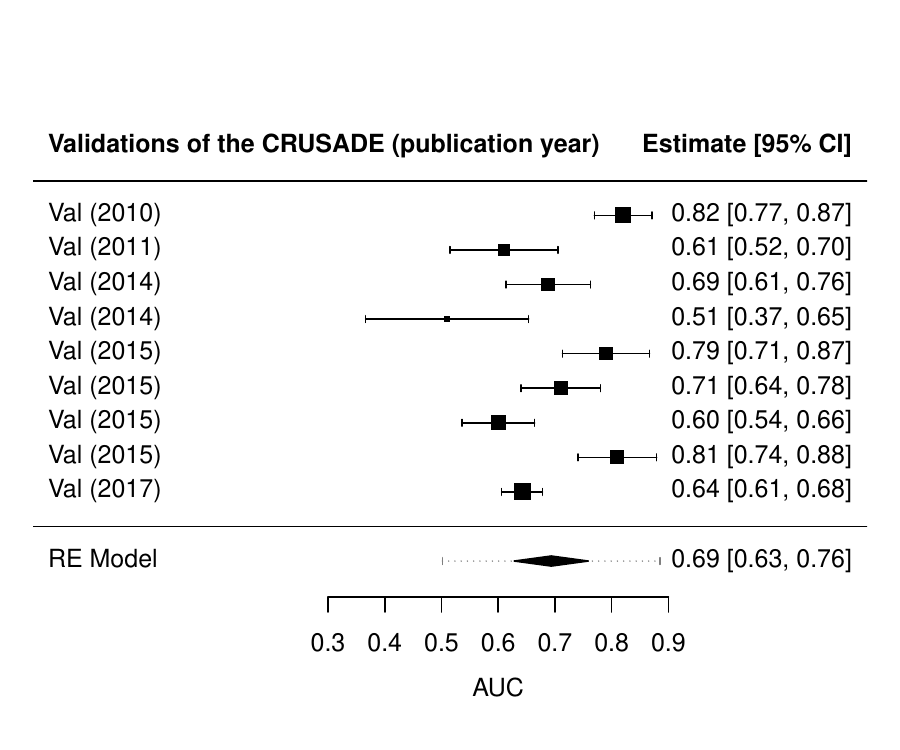} 

}

\caption{Forest plot and random effects meta-analysis of the AUC estimates of validations for the CRUSADE CPM. The black diamond is the 95\% confidence interval for the mean AUC across all validations. The dotted range represents the 95\% prediction interval for the true AUC in a new study. \label{fig:2}}\label{fig:unnamed-chunk-3}
\end{figure}

After the first external validation of the CRUSADE model, 8 more
validations were performed. We show the cumulative results in Figure
\ref{fig:2} as a forest plot. We used the R package metafor (Viechtbauer
2010) to do a standard random effects meta-analysis of all 9 external
validations. We estimate the pooled AUC to be 0.69 with 95\% confidence
interval from 0.63 to 0.76. Remarkably, this confidence interval
excludes the entire confidence interval after the first validation.

The large uncertainty about the pooled AUC is due to the large
heterogeneity between validation studies (Figure \ref{fig:2}). We
quantify this heterogeneity as the between-study standard deviation
\(\tau\), and in the case of the CRUSADE model we estimate
\(\tau =0.09\). The large heterogeneity may be due to many factors
including differences in population, standard of care, and variations in
predictor and outcome definitions and assessment (Van Calster et al.
2023), in addition to model misspecification.

In the case of meta-analysis of clinical trials, prediction intervals
for the effect of the treatment in a new study are recognized as
important (IntHout et al. 2016). Similarly, the 95\% prediction interval
for the AUC of a prediction model in a new setting is more relevant than
the 95\% confidence interval for the pooled AUC. We find that the
prediction interval based on the 9 external validations is centered at
0.69 and extends from 0.5 to 0.89---a range of discriminatory
performance that spans from useless to what most would consider very
good (De Hond, Steyerberg, and Van Calster 2022). Thus, even after 9
external validations, the performance in a new setting remains highly
uncertain.

\begin{figure}[ht]

{\centering \includegraphics[width=0.9\linewidth]{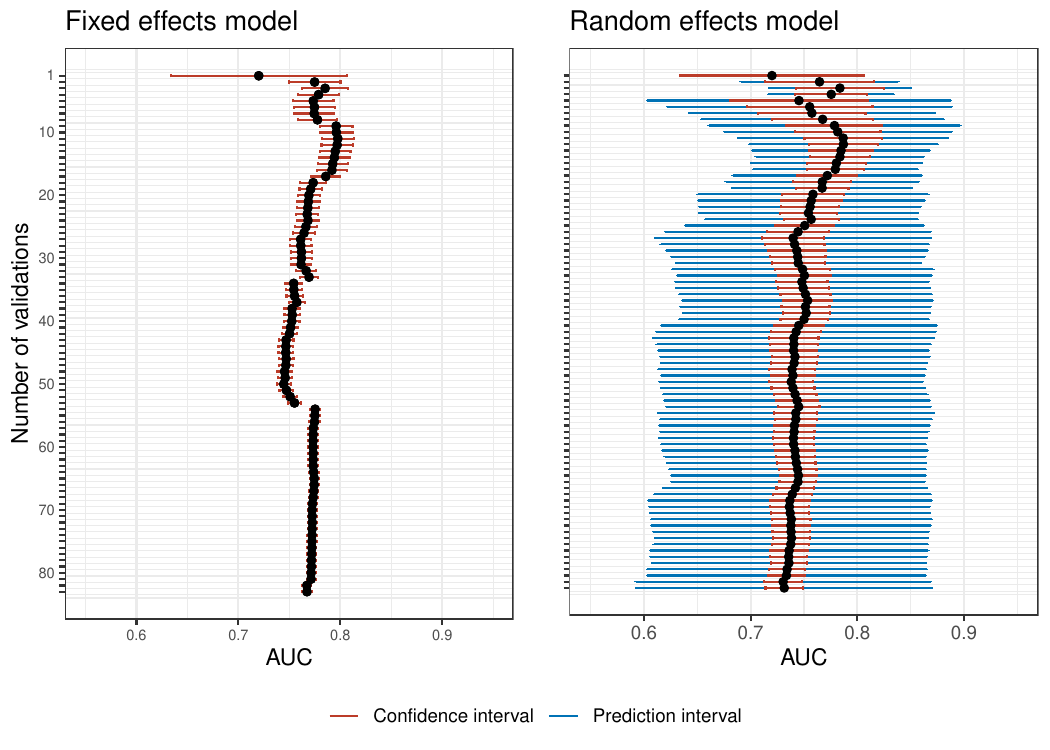} 

}

\caption{Cumulative confidence and prediction interval of fixed and random effects meta-analyses for the EuroScore model based on the first 1,2,3,...,83 validation studies.\label{fig:3}}\label{fig:unnamed-chunk-4}
\end{figure}

As a further illustration, consider the logistic EuroSCORE CPM for
patients undergoing major cardiac surgery (Roques et al. 2003). This
model has 83 external validations. In Figure \ref{fig:3} we show the
results of ``cumulative'' fixed and random effects meta-analyses. That
is, we show the 95\% confidence intervals for the mean AUC and the 95\%
prediction intervals for the AUC in a new setting based on the first
\(1,2,3,\dots,83\) validation studies. In the left panel, we show the
results of fixed effects meta-analyses. In that case, we assume that
\(\tau\) is zero and therefore the confidence interval for the pooled
AUC and the prediction interval for the AUC in a new study are equal.
After about 50 validations, the location of the intervals has stabilized
and their width has become negligible.

In the right panel, we show the results of random effects meta-analyses
where we used the REML method to estimate the heterogeneity \(\tau\).
When we have just one validation, it is not possible to estimate
\(\tau\) and it is set to zero. When we have few validations, the width
of the intervals vary considerably because the estimates of \(\tau\) are
very noisy. Eventually we see the width of the intervals stabilizing and
then gradually shrinking. While the width of the confidence interval
will tend to zero, the width of the prediction interval will not. In
fact, it will tend to \(2 \times 1.96 \times \tau\). Thus, no matter how
many validations have been done, there will always remain substantial
uncertainty about the AUC of the EuroSCORE model in a new setting.

It is obvious from Figure \ref{fig:3} that it is inappropriate to assume
that \(\tau\) is zero. This will lead to gross underestimation of the
uncertainty for the AUC. To make confidence intervals or prediction
intervals with the correct coverage, we need accurate estimates of
\(\tau\). Unfortunately, most CPMs have very few validation studies. Of
the CPMs included in our study, 239/469 (51\%) have only one external
validation. The median number of external validations is 2 with an IQR
from 1 to 3. Clearly, this is insufficient to estimate \(\tau\) with
good accuracy. Even worse, the usual methods (such as REML or the
well-known method of DerSimonian and Laird (DerSimonian and Laird
1986)) have a tendency to estimate \(\tau\) at zero. This happens because
the variation between the observed AUCs consists of within- and
between-study variation (heterogeneity). If the observed variation can
be explained by the within-study variation alone, then \(\tau\) will be
estimated at zero (Borenstein et al. 2010). As we will demonstrate, this
will often lead to severe undercoverage of confidence and prediction
intervals. This is the problem we want to address.

In the next section, we set up hierarchical (or multi-level) models to
study the 469 CPMs and their validations. In particular, we estimate the
distribution of \(\tau\) across the CPMs. We also estimate the
distribution of the pooled AUCs. Next, we implement two (empirical)
Bayesian models. The first has a flat prior for the average AUC, and an
informative prior for \(\tau\), and the second has informative priors
for both. We also have a ``poor man's'' Bayesian method where we set
\(\tau\) equal to a fixed (non-zero) value which can easily be done with
the metafor package (Viechtbauer 2010). To evaluate and compare the
frequentist and Bayesian methods, we use leave-one-study-out
cross-validation.

\hypertarget{methods}{%
\section{Methods}\label{methods}}

We use the observed AUC values of cardiovascular Clinical Prediction
Models (CPMs) from the Tufts PACE CPM Registry (Wessler et al. 2021).
This is a publicly available compilation of models predicting outcomes
for patients at risk for, or already having, cardiovascular disease. The
inclusion criteria of the registry require the CPM to predict a binary
cardiovascular outcome, presented in a way that enables patient risk
prediction. The search strategy considered CPMs that were developed and
published between 1990 and March 2015 case. Next, a SCOPUS citation
search on March 22, 2017, identified external validations of the CPMs,
defined as reports studying the same model in a new population. In
total, the registry has 1,382 CPMs and 2,030 external validations. Most
models are for patients with stroke (n= 97) and for patients undergoing
cardiac surgery (n=46). We selected CPMs with at least one external
validation and complete information. Thus our data consists of 469 CPMs
with 1,603 external validations (see the flowchart in the Appendix).

Since the validation AUCs are grouped within CPMs, we set up a
collection of random effects meta-analysis models (Whitehead and
Whitehead 1991). So, for the \(j\)-th validation AUC of the \(i\)-th
CPM, we assume:

\begin{equation}
\widehat{AUC}_{ij} \sim \mathcal{N}(AUC_{ij}, s_{ij}^2)
\end{equation}

\begin{equation}
AUC_{ij}\sim \mathcal{N}(AUC_i, \tau_i^2)
\end{equation} where \(i=1,2,\dots,n_j\), \(j=1,2,\dots,469\) and
\(s_{ij}\) denotes the standard error of the observed
\(\widehat{AUC}_{ij}\). Despite the fact that AUCs are bounded between 0
and 1, we believe the normal distribution is appropriate because the
observed values stay well away from the bounds (see Figure \ref{fig:1}).
As usual in meta-analyses, we will ignore the uncertainty about
\(s_{ij}\).

From the frequentist point of view, the \(AUC_i\) and \(\tau_i\) are
fixed parameters that are to be estimated. The defining feature of a
fixed effects meta-analysis is that \(\tau_i\) is assumed to be zero.
When \(\tau_i\) is not assumed to be zero, the metafor package has 12
different methods to estimate it (Viechtbauer 2010). Here, we use the
default REML method, but in the supplement we also consider the method
of Sidik and Jonkman (Sidik and Jonkman 2002) which tends to behave most
differently from REML among the remaining 11 methods. In our case,
however, the results turn out to be very similar to REML.

From the Bayesian perspective, we consider the \(AUC_i\) and \(\tau_i\)
to be random variables for which we need to specify prior distributions.
We will assume a normal distribution for the \(AUC_i\) and a lognormal
distribution for the \(\tau_i\):

\begin{equation}
AUC_{i}\sim \mathcal{N}(\mu_{AUC}, \sigma^2_{AUC})
\end{equation}

\begin{equation}
\log(\tau_{i}) \sim \mathcal{N}(\mu_\tau, \sigma_\tau^2)
\end{equation}

\noindent This implies that the mean and variance of the \(\tau_i\) are
\begin{equation}
\text{E}(\tau_i) = \exp\left( \mu_\tau + \frac{\sigma^2_\tau}{2} \right) \quad\text{and}\quad \text{Var}(\tau_i) = \left[ \exp(\sigma^2_\tau) - 1 \right] \exp(2\mu_\tau + \sigma^2_\tau).
\end{equation}

\noindent We use the method of maximum likelihood to estimate the 4
parameters of our model (\(\mu_{AUC}\), \(\sigma_{AUC}\), \(\mu_{\tau}\)
and \(\sigma_{\tau}\)). The likelihood does not have a closed form, so
we use the R-package rstan to do the computation (Stan Development
2023). This package provides an R interface to the Stan platform for
MCMC sampling to perform Bayesian inference. We specify uniform priors
for each of the 4 parameters, and then take their posterior modes as the
MLEs. In terms of the estimated model parameters, the estimated mean of
the \(\tau_i\) is \begin{equation}
\bar{\tau} = \exp\left( \hat{\mu}_\tau + \frac{\hat{\sigma}_\tau^2}{2} \right).
\end{equation}

Our main goal is to predict the AUC in a new setting, and to provide a
95\% prediction interval. We use the metafor package (Viechtbauer 2010)
to do 3 versions of frequentist meta-analyses:

\begin{enumerate}
  \item fixed effects model where we assume $\tau_i=0$,
  \item random effects where we estimate the $\tau_i$ with REML,
  \item random effects model where we assume $\tau_i=\bar{\tau}$.
\end{enumerate}

\noindent One could argue that the first and third model are actually
Bayesian models with extremely strong priors for the \(\tau_i\) and a
non-informative prior for the \(AUC_i\). We use the R-package baggr
(Wiecek and Meager 2022) to do two versions of (empirical) Bayesian
meta-analyses:

\begin{enumerate}
\setcounter{enumi}{3}
  \item Bayesian meta-analysis with a non-informative prior for $AUC_i$, and an informative prior for $\tau_i$,
  \item Bayesian meta-analysis with informative priors for both $AUC_i$ and $\tau_i$.
\end{enumerate}

To evaluate and compare the performance of these 5 methods, we use a
leave-one-study-out cross validation approach. We fix a number \(n\) of
validation studies (\(n=1,2,\dots,5\)) and then we use
(\(\widehat{AUC}_{i,1}\), \(s_{i,1}\)),\dots,(\(\widehat{AUC}_{i,n}\),
\(s_{i,n}\)) and \(s_{i,n+1}\) to predict \(\widehat{AUC}_{i,n+1}\). We
also form a 95\% prediction interval for \(\widehat{AUC}_{i,n+1}\). We
do this by forming the 95\% prediction interval for the true
\(AUC_{i,n+1}\), and then accounting for the sampling error. We show the
formulas in Table \ref{tab:1}.

We make sure that there at least \(n+1\) studies in the meta-analysis,
so that we can check how often the observed AUCs of the left-out studies
fall within the prediction interval. Hence, only CPMs with at least 2
validations are used in the cross-validation. If the coverage of the
observed AUCs is 95\% then we conclude that the coverage of the
prediction interval for the true AUC is also 95\%. Finally, we also
compute the root mean squared prediction error (RMSE) for the observed
AUC in a new study.

\begin{table}[!htbp]
\centering
\caption{Prediction intervals for the true and observed AUC in the next study. Based on the first $n$ studies, $\widehat{AUC}_{1:n}$ is the estimate of the pooled AUC, $s_{1:n}$ is the associated standard error, and $\hat{\tau}_{1:n}$ is the estimate of $\tau$. In the Bayesian approach, $AUC_{post}$ and $SD_{post}$ are the posterior mean and standard deviation of the pooled AUC. $s_{n+1}$ is the standard error of the observed AUC in the $(n+1)$-th study.}

\bigskip
\renewcommand{\arraystretch}{3}
\label{tab:1}
\begin{tabular}{l|l|l|l}
\textbf{Model} & \textbf{Description} & \textbf{Prediction interval for $AUC_{n+1}$} & \textbf{Prediction interval for $\widehat{AUC}_{n+1}$} \\
\hline
FE & $\tau = 0$  & $\widehat{AUC}_{1:n} \pm 1.96  \sqrt{s_{1:n}^2} $ & $\widehat{AUC}_{1:n} \pm 1.96 \sqrt{s_{1:n}^2 + s_{n+1}^2} $ \\ 
\hline

RE & Estimate $\tau$  & $\widehat{AUC}_{1:n} \pm 1.96 \sqrt{s_{1:n}^2 + \hat{\tau}_{1:n}^2} $ &    $\widehat{AUC}_{1:n} \pm 1.96 \sqrt{s_{1:n}^2 + \hat{\tau}_{1:n}^2 + s_{n+1}^2}$ \\ \hline
 
RE& $\tau = \bar{\tau}$ &  $\widehat{AUC}_{1:n} \pm 1.96 \sqrt{s_{1:n}^2 + \bar{\tau}^2} $ & $\widehat{AUC}_{1:n} \pm 1.96 \sqrt{s_{1:n}^2 + \bar{\tau}^2 + s_{n+1}^2} $\\ \hline

Bayes & Prior &    $\widehat{AUC}_{1:n} \pm 1.96 \sqrt{SD_{post}^2}$  &                  $\widehat{AUC}_{1:n} \pm 1.96 \sqrt{SD_{post}^2 + s_{n+1}^2}$
\end{tabular}
\end{table}

\hypertarget{results}{%
\section{Results}\label{results}}

Four parameters need to be estimated for our model, namely the mean and
standard deviation of the \(\tau_i\) and the mean and standard deviation
of the \(AUC_i\) (Table \ref{tab:2}). Note that we actually have two
variants; in the first variant we set the mean of the \(AUC_i\) to zero
and their standard deviation to a large value to obtain an essentially
flat or ``non-informative'' prior. The mean of the prior of \(\tau\) in
the first model is 0.055 with a standard deviation of 0.15. The mean and
standard deviation of \(\tau\) in model 2 are very similar at 0.057 and
0.12. For our fixed effects meta-analysis with non-zero heterogeneity,
we set \(\bar{\tau} = 0.055\).

\renewcommand{\arraystretch}{2}
\begin{table}[!htbp]
\centering
\caption{The estimated priors of the $\tau_i$ and $AUC_i$.}
\label{tab:2}
\begin{tabular}{l|l|l|l|l}
\textbf{Model} & $\mu_\tau$ & $\sigma_\tau$ & $\mu_{AUC}$ & $\sigma_{AUC}$ \\ 
\hline
Non-informative for $AUC_i$, informative for $\tau_i$  & -2.94  & 0.27 & 0  & 10  \\ 
\hline
Informative $AUC_i$ and $\tau_i$  & -2.89     & 0.21         & 0.73     & 0.07        
\end{tabular}
\end{table}

\begin{figure}[ht]

{\centering \includegraphics[width=0.9\linewidth]{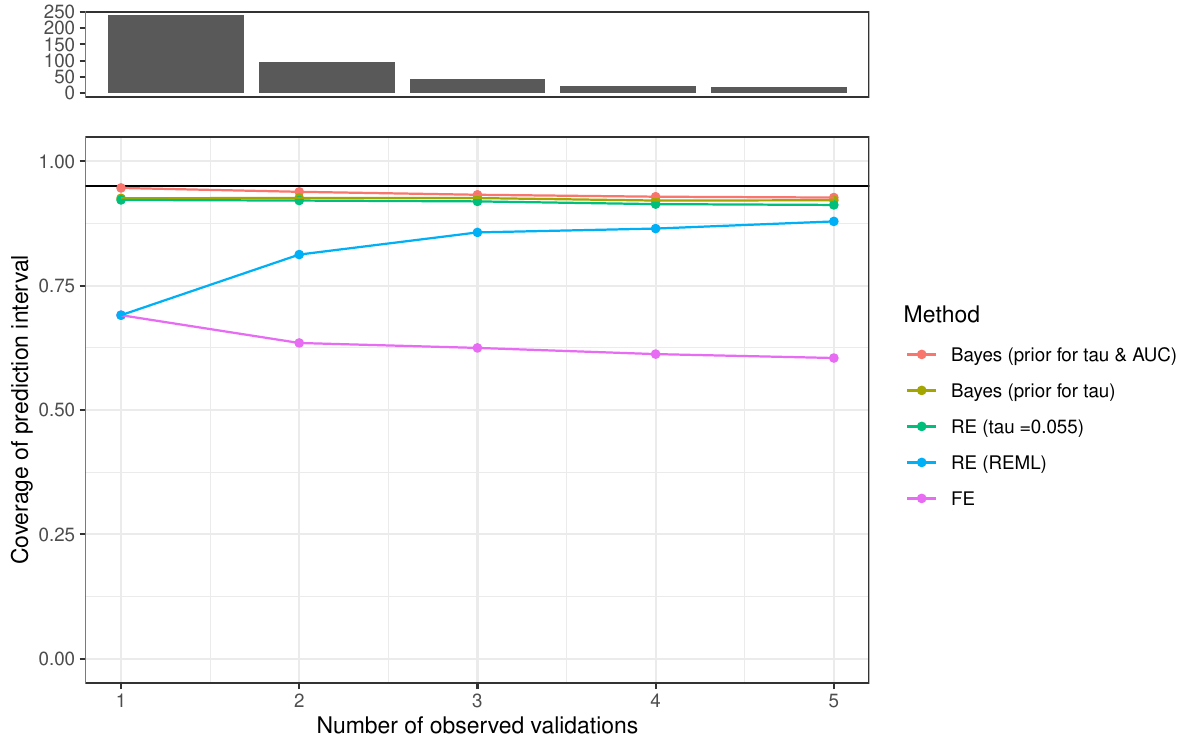} 

}

\caption{Top panel: the number of CPMs with exactly 1,2,...,5 external validations. Bottom panel: Coverage of the prediction intervals for the observed AUC in the next study.\label{fig:4}}\label{fig:unnamed-chunk-5}
\end{figure}

When the prediction intervals are based on only one study, both the
fixed effects model and the random effects model with REML estimation
can only set \(\tau\) equal to zero which results in severe
undercoverage (Figure \ref{fig:4}). The fixed effects model will
continue to undercover even when we base the prediction intervals on
more studies, but the coverage of the random effects model will increase
to the nominal level. When we base the prediction intervals on 5 or more
studies, the coverage of the random effects model becomes close to
nominal. However, only a small minority of CPMs (69/469, 15\%) have 5 or
more external validations. The two Bayesian models and the model where
we set \(\tau = 0.055\) always had near nominal coverage. The slight
undercoverage that remained may be expected from Wald type intervals
which ignore the uncertainty about the standard errors of the observed
AUCs.

\begin{figure}[ht]

{\centering \includegraphics[width=0.75\linewidth]{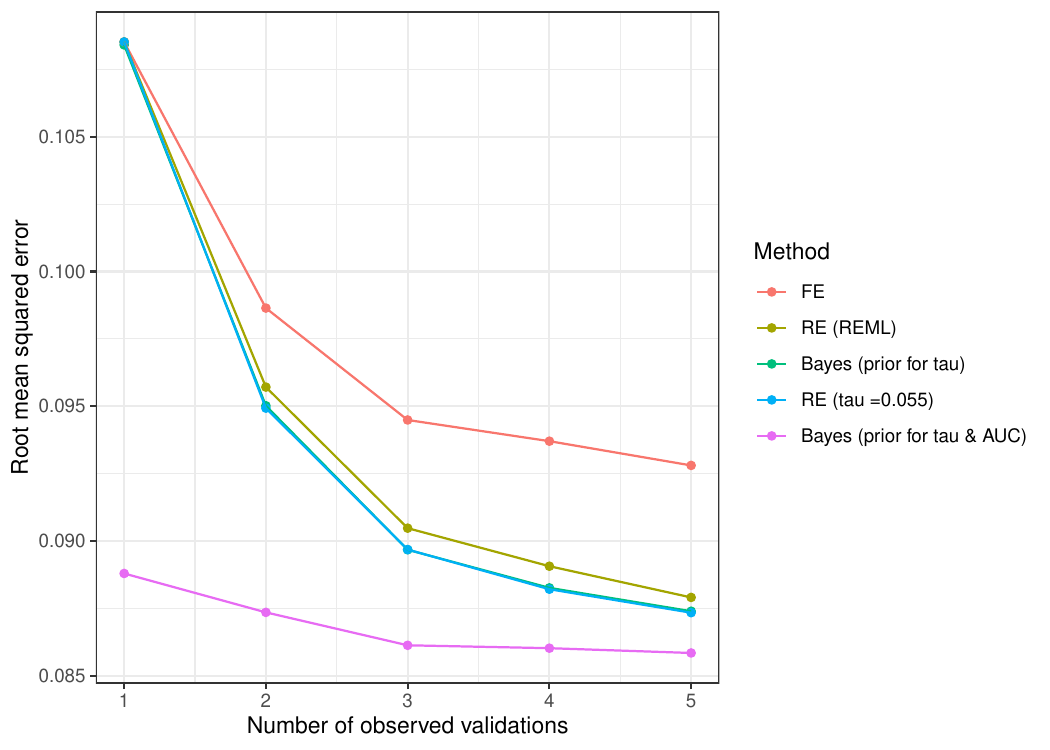} 

}

\caption{Root Mean Squared prediction Error for the observed AUC in the next study.\label{fig:5}}\label{fig:unnamed-chunk-6}
\end{figure}

We note the relatively poor performance of the fixed effects model,
which is due to the inefficient weighing of the individual studies
(Figure \ref{fig:5}). We also note the superior performance of the
Bayesian model with an informative prior for the AUC\_i which is due to
the shrinkage towards the overall average of the AUCs at 0.734. When we
use a single validation to predict the AUC in a new setting, the error
of Bayesian model is on average about 1 percentage point less than the
other methods. When we use more validations, this advantage decreases.

\hypertarget{discussion}{%
\section{Discussion}\label{discussion}}

We noted considerable heterogeneity among the external validations of
cardiovascular CPMs. We estimated that the standard deviation \(\tau\)
is about 0.05 on average with a standard deviation of 0.01.
Additionally, we estimated a normal distribution for the pooled AUCs
with a mean of 0.73 and a standard deviation of 0.07. Using these
distributions as an empirical prior substantially outperformed
frequentists methods of meta-analysis in terms of prediction accuracy
and coverage of the prediction interval for the next study. Especially
when there were few validation studies (fewer than 5), frequentist
methods showed severe undercoverage, while the empirical Bayes approach
was very close to nominal. Our study illustrates the usefulness of
empirical Bayes approaches for meta-analyses in general, where
estimation of heterogeneity is unreliable unless a large number of
studies is analyzed.

If \(\tau\) is 0.05, then the 95\% prediction interval for the AUC in a
new setting will have a width of at least +/- 0.1, no matter how many
validations have been done. In this sense, our findings verify the claim
of Van Calster et al.~(2023) that ``there is
no such thing as a validated prediction model''.

Obviously, external validations should be taken into account before
deployment of a CPM. However, most published CPMs have never been
externally validated (Siontis et al. 2015). When external validations
are done they do not provide a solid guarantee about the AUC in the next
study. Therefore the discriminatory performance in a new setting should
be monitored after deployment. While many researchers may understand the
AUC as an intrinsic measure of CPM quality, in fact AUC is an extrinsic
property of a CPM that emerges only when a model is applied to a
specific population.

There are two broad reasons for variation in AUC when transporting a
model from one setting to another: 1) differences in the heterogeneity
of the sample; 2) model misspecification. Regarding the first, more
heterogeneous populations will generally result in larger AUC values.
For example, at the extreme, a well specified 6-variable model will have
an AUC of exactly 0.5 if transported to a new population where each
patient has the same value for each of the 6 variables, even with fully
correct model specification.. This patient heterogeneity can be
quantified using various methods to measure the variance of predictions.
An intuitive summary is the standard deviation of the linear predictor
(Debray et al. 2015). Another important measure is the model-based
c-statistic, which is the c-statistic expected for a perfectly valid
model in the validation setting, based on the observed predictor values
(Debray et al. 2015). This benchmark for model performance could not be
calculated for our validations since we had no access to individual
patient data.

On the other hand, model invalidity reflects differences in the
associations of the predictor and outcome variables between the
derivation and validation samples. Such misspecification can arise for
many reasons, including changes in the population (particularly with
respect to the distribution of variables not included in the model that
may act as effect modifiers), changes with how data are collected or how
predictors or outcomes are defined, and changes in clinician and patient
behavior (Finlayson et al. 2021). Thus, the assumption of independence
of the outcome and data source (conditional on variables included in the
model) that undergirds prediction and transportability methods, is
commonly violated in actual practice. In a previous analysis, we
performed 158 validations of 108 published CPMs in the Tufts PACE
registry (Gulati et al. 2022). We used publicly available data from
randomized controlled trials for validation, where we expect less
heterogeneity than in less selected observational data sources as
typically used for development of CPMs. We found that the AUC differed
substantially between model derivation (0.76 {[}interquartile range
0.73-0.78{]}) and validation (0.64 {[}interquartile range 0.60-0.67{]}).
Indeed, approximately half of this decrease could be accounted for by
the narrower case-mix (less heterogeneity) in the validation samples;
the remainder could be attributed to model misspecification.

Moreover, it can be argued that the AUC does not provide the most
pertinent information about the usefulness of the CPM. The AUC is a
measure of discrimination across all possible cut-offs and as such it is
not directly meaningful when a particular cut-off is used in clinical
practice to support decision making. Decision-analytic summary measures
such as Net Benefit quantify clinical usefulness better (Vickers, Van
Calster, and Steyerberg 2016). Net Benefit depends on discrimination
(higher with higher AUC) and calibration (highest with correct
calibration at the decision threshold). Moreover, the clinical context
is important, with higher Net Benefit if the decision threshold is in
the middle of the risk distribution. Further work is necessary on
quantifying calibration across validations of CPMs. A natural starting
point is to quantify heterogeneity in summary measures for calibration
in the large, where poor validity is commonly observed (Van Calster et
al. 2019).

We conclude that if we want to predict the AUC in a new setting, then
the uncertainty due to the heterogeneity among the validations is at
least comparable the sampling uncertainty. The proposed empirical Bayes
approach merits further implementation to properly address uncertainty
in CPM performance.

\hypertarget{data-and-code}{%
\section{Data and Code}\label{data-and-code}}

The Tufts PACE CPM Registry is publicly available at
\texttt{www.pacecpmregistry.org}. supplement.

\newpage

\hypertarget{appendix}{%
\section{Appendix}\label{appendix}}

\begin{figure}[ht]

{\centering \includegraphics[width=0.9\linewidth]{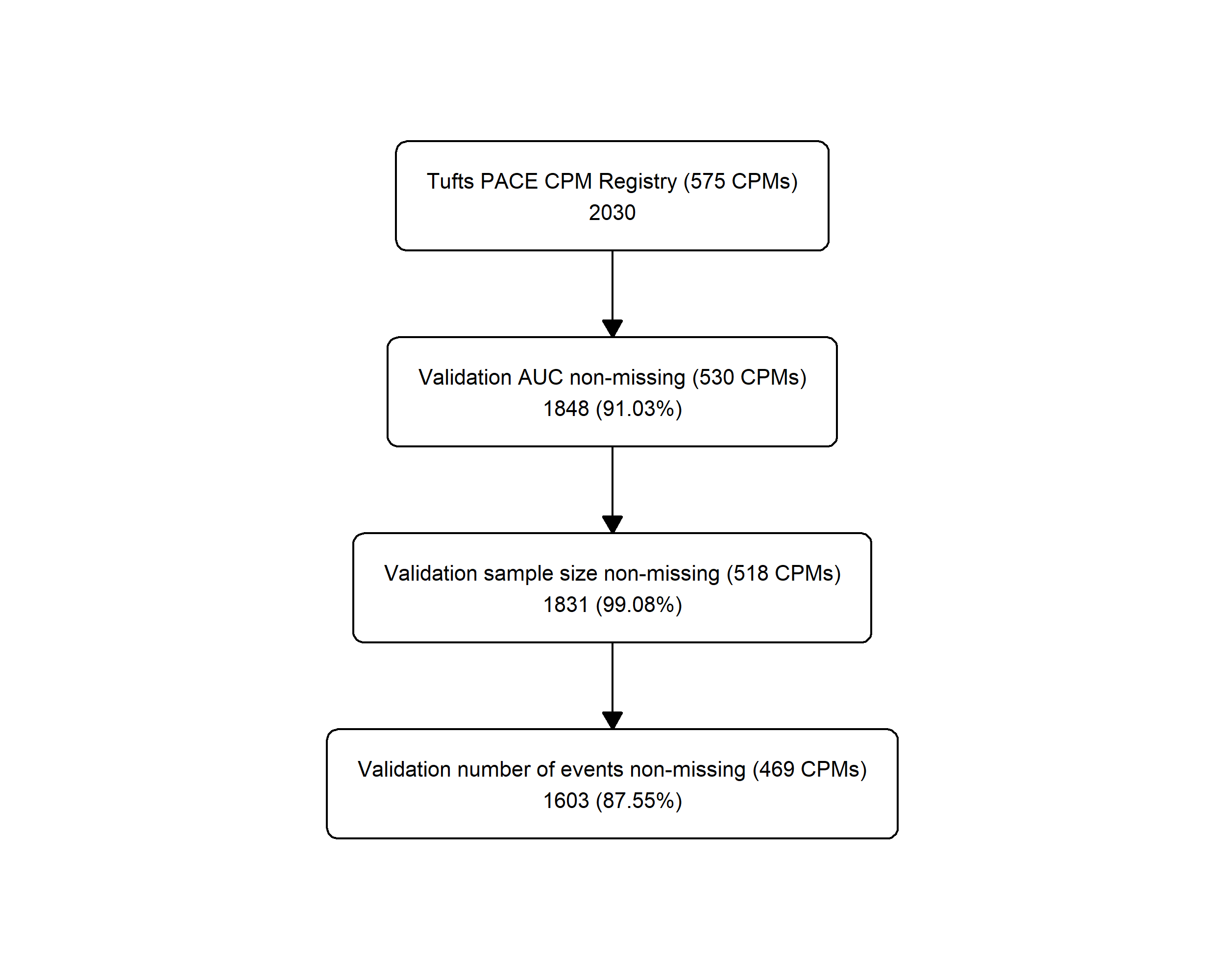} 

}

\caption{Flowchart of data filtering.\label{fig:6}}\label{fig:unnamed-chunk-7}
\end{figure}

The data selection is shown in the flowchart (Figure \ref{fig:6}). At
the start there are a total of 2,030 validations of 575 CPMs. After
filtering we have 1,603 validations from 469 CPMs.

\newpage

\section*{References}

\hypertarget{refs}{}
\begin{CSLReferences}{1}{0}
\leavevmode\vadjust pre{\hypertarget{ref-abu-assi_evaluating_2010}{}}%
Abu-Assi, Emad, José María Gracía-Acuña, Ignacio Ferreira-González,
Carlos Peña-Gil, Pilar Gayoso-Diz, and José Ramón González-Juanatey.
2010. {``Evaluating the Performance of the Can Rapid Risk Stratification
of Unstable Angina Patients Suppress Adverse Outcomes with Early
Implementation of the {ACC}/{AHA} Guidelines ({CRUSADE}) Bleeding Score
in a Contemporary Spanish Cohort of Patients with Non--{ST}-Segment
Elevation Acute Myocardial Infarction.''} \emph{Circulation} 121 (22):
2419--26.
\url{https://www.ahajournals.org/doi/10.1161/CIRCULATIONAHA.109.925594}.

\leavevmode\vadjust pre{\hypertarget{ref-altman_what_2000}{}}%
Altman, Douglas G., and Patrick Royston. 2000. {``What Do We Mean by
Validating a Prognostic Model?''} \emph{Statistics in Medicine} 19 (4):
453--73.

\leavevmode\vadjust pre{\hypertarget{ref-borenstein_basic_2010}{}}%
Borenstein, Michael, Larry V. Hedges, Julian P. T. Higgins, and Hannah
R. Rothstein. 2010. {``A Basic Introduction to Fixed-Effect and
Random-Effects Models for Meta-Analysis.''} \emph{Research Synthesis
Methods} 1 (2): 97--111.
\url{https://onlinelibrary.wiley.com/doi/10.1002/jrsm.12}.

\leavevmode\vadjust pre{\hypertarget{ref-hond_interpreting_2022}{}}%
De Hond, Anne A. H., Ewout W. Steyerberg, and Ben Van Calster. 2022.
{``Interpreting Area Under the Receiver Operating Characteristic
Curve.''} \emph{The Lancet Digital Health} 4 (12): e853--55.
\url{https://www.thelancet.com/journals/landig/article/PIIS2589-7500(22)00188-1/fulltext}.

\leavevmode\vadjust pre{\hypertarget{ref-debray_new_2015}{}}%
Debray, Thomas P. A., Yvonne Vergouwe, Hendrik Koffijberg, Daan Nieboer,
Ewout W. Steyerberg, and Karel G. M. Moons. 2015. {``A New Framework to
Enhance the Interpretation of External Validation Studies of Clinical
Prediction Models.''} \emph{Journal of Clinical Epidemiology} 68 (3):
279--89.

\leavevmode\vadjust pre{\hypertarget{ref-dersimonian_meta-analysis_1986}{}}%
DerSimonian, Rebecca, and Nan Laird. 1986. {``Meta-Analysis in Clinical
Trials.''} \emph{Controlled Clinical Trials} 7 (3): 177--88.
\url{https://www.sciencedirect.com/science/article/pii/0197245686900462}.

\leavevmode\vadjust pre{\hypertarget{ref-finlayson_clinician_2021}{}}%
Finlayson, Samuel G., Adarsh Subbaswamy, Karandeep Singh, John Bowers,
Annabel Kupke, Jonathan Zittrain, Isaac S. Kohane, and Suchi Saria.
2021. {``The Clinician and Dataset Shift in Artificial Intelligence.''}
\emph{New England Journal of Medicine} 385 (3): 283--86.
\url{http://www.nejm.org/doi/10.1056/NEJMc2104626}.

\leavevmode\vadjust pre{\hypertarget{ref-gulati_generalizability_2022}{}}%
Gulati, Gaurav, Jenica Upshaw, Benjamin S. Wessler, Riley J. Brazil,
Jason Nelson, David Van Klaveren, Christine M. Lundquist, et al. 2022.
{``Generalizability of Cardiovascular Disease Clinical Prediction
Models: 158 Independent External Validations of 104 Unique Models.''}
\emph{Circulation: Cardiovascular Quality and Outcomes} 15 (4).
\url{https://www.ahajournals.org/doi/10.1161/CIRCOUTCOMES.121.008487}.

\leavevmode\vadjust pre{\hypertarget{ref-harrell_describing_2015}{}}%
Harrell, Frank E. 2015. \emph{Regression Modeling Strategies}. Cham:
Springer International Publishing.

\leavevmode\vadjust pre{\hypertarget{ref-inthout_plea_2016}{}}%
IntHout, Joanna, John PA Ioannidis, Maroeska M. Rovers, and Jelle J.
Goeman. 2016. {``Plea for Routinely Presenting Prediction Intervals in
Meta-Analysis.''} \emph{{BMJ} Open} 6 (7): e010247.
\url{https://bmjopen.bmj.com/content/6/7/e010247.abstract}.

\leavevmode\vadjust pre{\hypertarget{ref-justice_assessing_1999}{}}%
Justice, Amy C. 1999. {``Assessing the Generalizability of Prognostic
Information.''} \emph{Annals of Internal Medicine} 130 (6): 515.
\url{http://annals.org/article.aspx?doi=10.7326/0003-4819-130-6-199903160-00016}.

\leavevmode\vadjust pre{\hypertarget{ref-roques_logistic_2003}{}}%
Roques, Frangois, Philippe Michel, A. R. Goldstone, and S. A. M. Nashef.
2003. {``The Logistic Euroscore.''} \emph{European Heart Journal} 24
(9): 882--83.
\url{https://academic.oup.com/eurheartj/article-abstract/24/9/882/2733949}.

\leavevmode\vadjust pre{\hypertarget{ref-sidik_simple_2002}{}}%
Sidik, Kurex, and Jeffrey N. Jonkman. 2002. {``A Simple Confidence
Interval for Meta-Analysis.''} \emph{Statistics in Medicine} 21 (21):
3153--59.
\url{https://onlinelibrary.wiley.com/doi/abs/10.1002/sim.1262}.

\leavevmode\vadjust pre{\hypertarget{ref-siontis_external_2015}{}}%
Siontis, George C. M., Ioanna Tzoulaki, Peter J. Castaldi, and John P.
A. Ioannidis. 2015. {``External Validation of New Risk Prediction Models
Is Infrequent and Reveals Worse Prognostic Discrimination.''}
\emph{Journal of Clinical Epidemiology} 68 (1): 25--34.

\leavevmode\vadjust pre{\hypertarget{ref-stan_development_rstan_2023}{}}%
Stan Development, Team. 2023. {``{RStan}: The r Interface to Stan.''}
\emph{R Package Version 2.32.3.}

\leavevmode\vadjust pre{\hypertarget{ref-steyerberg_applications_2009}{}}%
Steyerberg, E. W. 2009. \emph{Clinical Prediction Models}. New York,
{NY}: Springer New York.

\leavevmode\vadjust pre{\hypertarget{ref-steyerberg_prediction_2016}{}}%
Steyerberg, Ewout W., and Frank E. Harrell. 2016. {``Prediction Models
Need Appropriate Internal, Internal-External, and External
Validation.''} \emph{Journal of Clinical Epidemiology} 69: 245.
\url{https://www.ncbi.nlm.nih.gov/pmc/articles/PMC5578404/}.

\leavevmode\vadjust pre{\hypertarget{ref-subherwal_baseline_2009}{}}%
Subherwal, Sumeet, Richard G. Bach, Anita Y. Chen, Brian F. Gage, Sunil
V. Rao, L. Kristin Newby, Tracy Y. Wang, et al. 2009. {``Baseline Risk
of Major Bleeding in Non--{ST}-Segment--Elevation Myocardial Infarction:
The {CRUSADE} Bleeding Score.''} \emph{Circulation} 119 (14): 1873--82.
\url{https://www.ahajournals.org/doi/10.1161/CIRCULATIONAHA.108.828541}.

\leavevmode\vadjust pre{\hypertarget{ref-van_calster_calibration_2019}{}}%
Van Calster, Ben, David J. McLernon, Maarten Van Smeden, Laure Wynants,
and Ewout W. Steyerberg. 2019. {``Calibration: The Achilles Heel of
Predictive Analytics.''} \emph{{BMC} Medicine} 17 (1).
\url{https://bmcmedicine.biomedcentral.com/articles/10.1186/s12916-019-1466-7}.

\leavevmode\vadjust pre{\hypertarget{ref-van_calster_there_2023}{}}%
Van Calster, Ben, Ewout W. Steyerberg, Laure Wynants, and Maarten Van
Smeden. 2023. {``There Is No Such Thing as a Valiyeard Prediction
Model.''} \emph{{BMC} Medicine} 21 (1): 70.
\url{https://bmcmedicine.biomedcentral.com/articles/10.1186/s12916-023-02779-w}.

\leavevmode\vadjust pre{\hypertarget{ref-vickers_net_2016}{}}%
Vickers, Andrew J, Ben Van Calster, and Ewout W Steyerberg. 2016. {``Net
Benefit Approaches to the Evaluation of Prediction Models, Molecular
Markers, and Diagnostic Tests.''} \emph{{BMJ}}, i6.
\url{https://www.bmj.com/lookup/doi/10.1136/bmj.i6}.

\leavevmode\vadjust pre{\hypertarget{ref-viechtbauer_conducting_2010}{}}%
Viechtbauer, Wolfgang. 2010. {``Conducting Meta-Analyses in r with the
Metafor Package.''} \emph{Journal of Statistical Software} 36 (3):
1--48.

\leavevmode\vadjust pre{\hypertarget{ref-wessler_external_2021}{}}%
Wessler, Benjamin S., Jason Nelson, Jinny G. Park, Hannah McGinnes,
Gaurav Gulati, Riley Brazil, Ben Van Calster, et al. 2021. {``External
Validations of Cardiovascular Clinical Prediction Models: A Large-Scale
Review of the Literature.''} \emph{Circulation: Cardiovascular Quality
and Outcomes} 14 (8): e007858.
\url{https://doi.org/10.1161/CIRCOUTCOMES.121.007858}.

\leavevmode\vadjust pre{\hypertarget{ref-whitehead_general_1991}{}}%
Whitehead, Anne, and John Whitehead. 1991. {``A General Parametric
Approach to the Meta‐analysis of Randomized Clinical Trials.''}
\emph{Statistics in Medicine} 10 (11): 1665--77.
\url{https://onlinelibrary.wiley.com/doi/10.1002/sim.4780101105}.

\leavevmode\vadjust pre{\hypertarget{ref-wiecek_baggr_2022}{}}%
Wiecek, Witold, and Rachael Meager. 2022. {``Baggr: Bayesian Aggregate
Treatment Effects.''} \emph{R Package Version 0.7.6} 18.

\end{CSLReferences}

\bibliographystyle{unsrt}
\bibliography{references.bib}

\end{document}